\documentclass[sn-basic]{sn-jnl}% Basic Springer Nature Reference Style/Chemistry Reference Style
\usepackage{graphicx}%
\usepackage{multirow}%
\usepackage{amsmath,amssymb,amsfonts}%
\usepackage{amsthm}%
\usepackage{mathrsfs}%
\usepackage[title]{appendix}%
\usepackage{xcolor}%
\usepackage{textcomp}%
\usepackage{manyfoot}%
\usepackage{booktabs}%
\usepackage{algorithm}%
\usepackage{algorithmicx}%
\usepackage{algpseudocode}%
\usepackage{listings}%
\usepackage{natbib}
\usepackage{lmodern}
\usepackage{float}
\begin{document}

\title[Angio-Diff: Learning a Self-Supervised Adversarial Diffusion Model for Angiographic Geometry Generation]{Angio-Diff: Learning a Self-Supervised Adversarial Diffusion Model for Angiographic Geometry Generation}

%%=============================================================%%
%% GivenName	-> \fnm{Joergen W.}
%% Particle	-> \spfx{van der} -> surname prefix
%% FamilyName	-> \sur{Ploeg}
%% Suffix	-> \sfx{IV}
%% \author*[1,2]{\fnm{Joergen W.} \spfx{van der} \sur{Ploeg} 
%%  \sfx{IV}}\email{iauthor@gmail.com}
%%=============================================================%%

\author[1]{\fnm{Zhifeng} \sur{Wang}}\email{zhifengwang@nudt.edu.cn}
% \equalcont{These authors contributed equally to this work.}

\author*[1]{\fnm{Renjiao} \sur{Yi}}\email{yirenjiao@nudt.edu.cn}
% \equalcont{These authors contributed equally to this work.}

\author[1]{\fnm{Xin} \sur{Wen}}\email{wenxin21@nudt.edu.cn}

\author[1]{\fnm{Chenyang} \sur{Zhu}}\email{zhuchenyang07@nudt.edu.cn}

\author[1]{\fnm{Kai} \sur{Xu}}\email{kevin.kai.xu@gmail.com}

\author*[2]{\fnm{Kunlun} \sur{He}}\email{kunlunhe@plagh.org}
% \author*[1]{\fnm{First} \sur{Author}}\email{Kai Xu}

\affil*[1]{\orgdiv{College of Computer Science}, \orgname{National University of Defense Technology}, \orgaddress{\street{No. 109, Deya Road}, \city{Changsha}, \postcode{410073}, \state{Hunan}, \country{China}}}

\affil*[2]{\orgdiv{Medical Big Data Research Center}, \orgname{Chinese PLA General Hospital}, \orgaddress{\city{Beijing}, \postcode{100853}, \country{China}}}

%%==================================%%
%% Sample for unstructured abstract %%
%%==================================%%

\abstract{Vascular diseases pose a significant threat to human health, with X-ray angiography established as the gold standard for diagnosis, allowing for detailed observation of blood vessels. However, angiographic X-rays expose personnel and patients to higher radiation levels than non-angiographic X-rays, which are unwanted. Thus, modality translation from non-angiographic to angiographic X-rays is desirable. Data-driven deep approaches are hindered by the lack of paired large-scale X-ray angiography datasets. While making high-quality vascular angiography synthesis crucial, it remains challenging. We find that current medical image synthesis primarily operates at pixel level and struggles to adapt to the complex geometric structure of blood vessels, resulting in unsatisfactory quality of blood vessel image synthesis, such as disconnections or unnatural curvatures. To overcome this issue, we propose a self-supervised method via diffusion models to transform non-angiographic X-rays into angiographic X-rays, mitigating data shortages for data-driven approaches. Our model comprises a diffusion model that learns the distribution of vascular data from diffusion latent, a generator for vessel synthesis, and a mask-based adversarial module. 
To enhance geometric accuracy, we propose a parametric vascular model to fit the shape and distribution of blood vessels. The proposed method contributes a pipeline and a synthetic dataset for X-ray angiography. 
We conducted extensive comparative and ablation experiments to evaluate the Angio-Diff. The results demonstrate that our method achieves state-of-the-art performance in synthetic angiography image quality and more accurately synthesizes the geometric structure of blood vessels. The code is available at \href{https://github.com/zfw-cv/AngioDiff}{\texttt{https://github.com/zfw-cv/AngioDiff}}.}


%%\pacs[JEL Classification]{D8, H51}

%%\pacs[MSC Classification]{35A01, 65L10, 65L12, 65L20, 65L70}

\maketitle

\section{Introduction}
\label{sec:intro}

Vascular disease stands as one of the paramount health afflictions worldwide. To diagnose them, X-ray angiography provides good observations of vascular structures, giving detailed insights into vascular narrowings or obstructions. X-ray angiography has become the gold standard for diagnosing coronary vascular disease. Consequently, healthcare professionals typically rely on angiographic imagery to characterize issues such as vascular wall calcification and stenosis. 
However, this diagnostic method takes a longer time than non-angiographic X-rays. It is an invasive examination using contrast agents, which may cause certain allergic reactions and a larger amount of radiation exposure associated with contrast agent usage. 
Therefore, exploring a non-invasive vascular angiography approach holds significant clinical diagnostic implications. This paper proposes a self-supervised adversarial diffusion method to synthesize angiographic X-rays with precise vascular structures from non-angiographic X-rays.
\begin{figure}[ht]
  \centering
  \includegraphics[width=\linewidth]{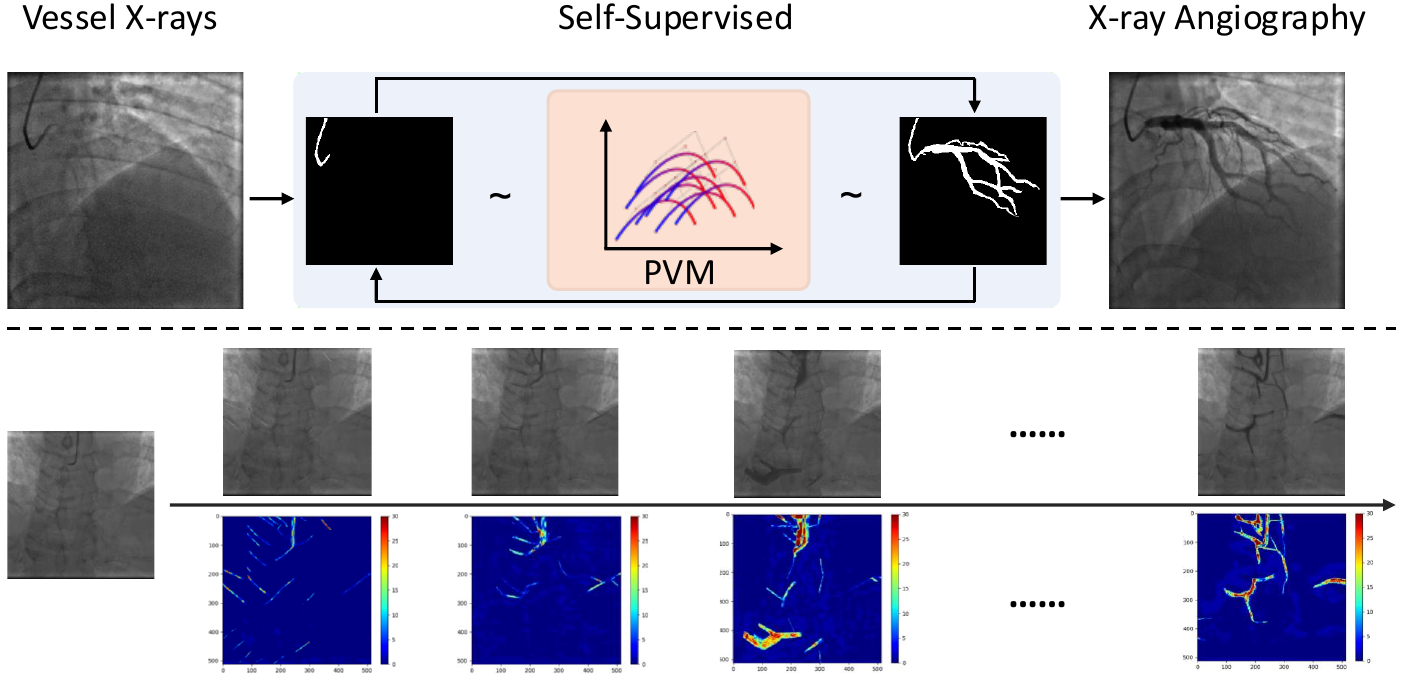}
  \caption{Top: Non-angiographic X-rays are translated into synthetic angiograms via an adversarial diffusion model guided by a parametric vascular model(PVM). Bottom: The synthesis process is demonstrated, with heatmaps illustrating vascular geometries' progressive generation and refinement.}
  \label{fig:teaser}
  % \vspace{-0.6cm}
\end{figure}

Synthesizing high-quality vascular angiography is a challenging problem. First of all, the current methods of synthetic vascular angiography hardly consider the geometric structure of blood vessels, or the vessel geometry is not well adapted, resulting in unnatural vascular synthesis. Secondly, it is difficult to obtain paired image data, and the cost of manual professional annotation of data is very high. 

To simulate high-quality vascular angiography, Ma et al.~\cite{ma2021self} introduced the geometric structure of blood vessels, designing a fractal synthesis module based on rectangular blocks that can generate a variety of vascular shapes. However, this method lacks accurate descriptions of the geometric characteristics of the blood vessels. Differently, Nagayama et al.~\cite{nagayama2023improving} developed a super-resolution reconstruction~\cite{peng2024towards} technique to enhance the image quality of Cardiac Computed Tomography Angiography (CCTA). This approach focuses on pixel-level image enhancement within a single modality and has achieved better medical image enhancement. However, it is not suitable for synthesizing images from other modalities. Leveraging the continuous and smooth geometric characteristics of the vasculature, we propose a parametric vascular model based on cubic Bézier curves~\cite{yan2016adjustable} to simulate the geometric morphology of blood vessels. This module imposes geometric constraints on the task of vascular synthesis, enabling the generation of high-quality angiographic X-rays, as shown in Figure~\ref{fig:teaser}.

While much research has focused on Computed Tomography Angiography (CTA) synthesis, angiographic X-rays (XCA), as an interventional imaging method, provide higher resolution vascular images than CTA and allow direct observation of the internal conditions of blood vessels. This makes XCA advantageous for diagnosing vascular stenosis and calcification. However, collecting and annotating paired XCA datasets for model training remains labor-intensive and costly.

To address this challenge, we propose a self-supervised pipeline to synthesize angiographic X-rays from non-angiographic X-rays without requiring paired training data. The method leverages a diffusion model to learn the data distribution of angiographic X-rays and generates pseudo-angiographic image pairs during adversarial learning. Additionally, a parametric vascular model based on Bézier curves~\cite{wen2005bezier} is introduced, where control points are adjusted according to vessel distribution gradients to accurately capture vascular geometry. This approach facilitates the generation of large quantities of synthetic paired data, providing a robust solution for data-driven learning-based applications. Empirical results demonstrate the effectiveness of the proposed method in improving the precision of vascular geometry modeling.

The main contributions of this paper are:
\begin{itemize}
	\item Angio-Diff is a self-supervised adversarial diffusion model designed for generating angiographic vascular geometries. It effectively synthesizes paired non-angiographic and angiographic X-ray data.  
	\item We design a novel parametric vascular model tailored to model vascular geometries, which has achieved state-of-the-art performance on pertinent datasets.
 	\item A large angiographic dataset, SynthXCA, was synthesized using our method and achieves state-of-the-art performance in both angiographic quality and vascular geometry. This dataset will be publicly released, with sample data available in the supplementary materials. 
\end{itemize}

% \vspace{-0.2cm}
\section{Related Work}
\label{sec:formatting}
%-------------------------------------------------------------------------
\subsection{Cross-modality medical image synthesis}\label{cross_medical}
In the field of medical image processing, cross-modal image synthesis aims to convert images from different modalities to facilitate information sharing and data augmentation across various imaging techniques.
Early methods relied on feature extractors and image registration~\cite{peng2024efficient} for achieving cross-modal image synthesis. Chen et al.~\cite{chen2017cross} introduced a multi-channel image registration method using image synthesis to generate proxy images for missing modalities. Isola et al.~\cite{isola2017image} proposed a conditional generative adversarial network to achieve image-to-image conversion.
However, these methods are limited by feature representation and matching accuracy, which hinders their effectiveness in complex medical image conversion tasks. In contrast, deep learning-based approaches, which directly learn the mapping between cross-modal images from data, have recently demonstrated superior performance. Among these, Generative Adversarial Networks (GANs) are widely used for cross-modal medical image synthesis. For instance, 
~\cite{zhu2017unpaired,azzam2020ktransgan,zhang2023minimalgan,cheema2024novel,mahmood2018unsupervised,sariyildiz2023fake,he2024dual,song2025uni,song2023alias} introduce a generator network and a discriminator network to generate high-quality synthetic images while enhancing their fidelity through adversarial training.~\cite{li2023zero,kazerouni2023diffusion,guo2024image} utilizing diffusion models for unsupervised image translation, both approaches leverage frequency guidance or adversarial learning techniques to achieve high-quality medical image conversion.

Although existing pixel-based modality conversion methods and generative approaches based on GANs or diffusion models have made notable progress, they largely overlook the inherent geometric structure of blood vessels. However, in clinical practice, vascular geometry plays a critical role in diagnosing many diseases. To address this gap, this paper proposes a parameterized vascular model that explicitly captures vascular geometry, enabling the synthesis of anatomically realistic vessels in angiographic images.

%-------------------------------------------------------------------------
% \vspace{-0.8cm}
\subsection{Vascular synthesis}
Vascular synthesis technology is an advanced computer graphics method that simulates realistic vascular networks through algorithmic generation. These techniques are typically grounded in biological principles and mathematical models to ensure that the generated vascular structures are both natural and physiologically plausible. At the core of vascular synthesis technology lies the creation of a multi-layered, intricately branching network, mimicking the distribution and morphology of vasculature in real biological organisms. Galarreta. et al.~\cite{galarreta2013three} utilized stochastic L-systems to generate three-dimensional synthetic vascular models, employing computational algorithms to simulate natural growth processes for creating more realistic and complex vascular networks, holding potential applications in tissue engineering and regenerative medicine fields. Wolterink et al.~\cite{wolterink2018blood} introduced a deep generative model based on Generative Adversarial Networks (GANs) for synthesizing vascular geometric structures, particularly suited for coronary arteries in cardiac CT angiography (CCTA) by training Wasserstein GANs, including both generator and discriminator networks, to achieve synthesis and evaluation of vascular models in medical image analysis applications. There are also many self-supervised methods for image synthesis~\cite{krishnan2022self,wang2024cardiovascular,wang2022self,shi2024self,cao2024semi,kim2022diffusion,hu2024sali}. Feldman et al.~\cite{feldman2023vesselvae} proposed a data-driven generative framework utilizing recursive variational autoencoders to synthesize three-dimensional vascular geometric structures. 

These data-driven methods rely on high-quality training data, which is often expensive. In contrast, this study introduces a parametric vascular model that guides the synthesis of high-quality vessels geometrically, enabling the generation of synthetic data to support downstream medical tasks.

\subsection{Diffusion models in medical imaging}
Stable Diffusion~\cite{gao2023corediff,ozbey2023unsupervised,yu2024diff,wang2025vastsd} have been successfully applied to medical image analysis, image segmentation, and other tasks. For instance, Khader et al.~\cite{khader2023denoising} utilize diffusion probabilistic models to generate high-quality 3D MRI and CT images. Similarly, Kidder~\cite{kidder2023advanced} has employed stable diffusion techniques to generate various medical images, including MRI, chest and lung X-rays, and contrast-enhanced spectral mammography (CESM) images. Furthermore, Ozbey et al.~\cite{ozbey2023unsupervised} and Jiang et al.~\cite{jiang2023cola} employ a conditional diffusion process to map noise and source images to target images. This approach achieves efficient and high-fidelity intermodal conversion.

Although diffusion has many applications in the medical field, there is currently a lack of research specifically focused on synthetic tasks for blood vessels. Additionally, from a cross-domain synthesis perspective, there is a notable absence of work on the transformation from non-angiographic X-rays to XCA.

\section{Method}
\begin{figure*}
  \centering
  \includegraphics[width=0.95\textwidth]{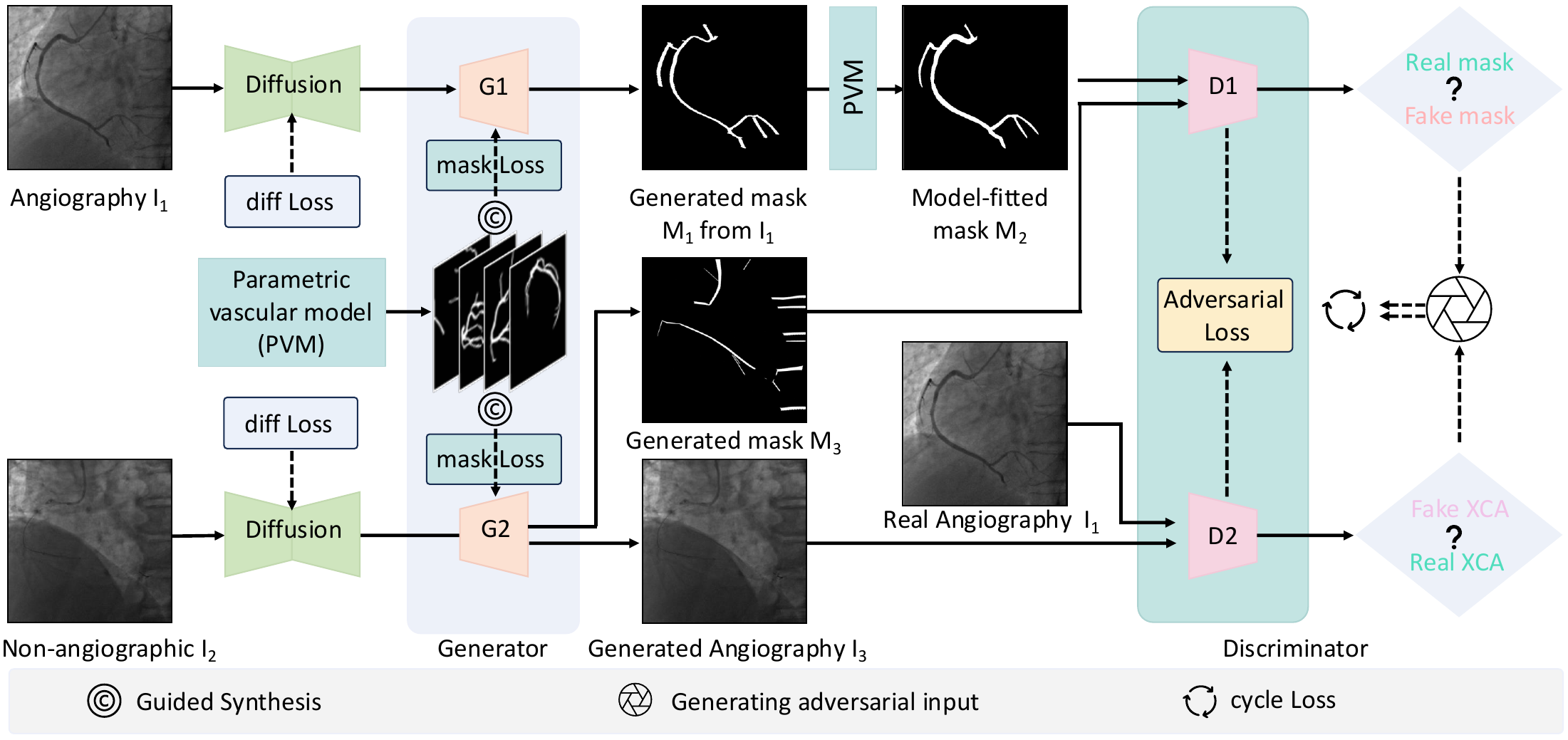} 
  \caption{Pipeline of self-supervised vascular angiography synthesis via adversarial diffusion models. The diffusion model takes an unpaired vascular angiography image $I_1$ and a non-angiographic X-ray image $I_2$ as inputs, generating a map of noise data distribution. This map is then fed into the generator. Moreover, a parametric vascular model produces an initial vascular mask as semantic input for the generator. Generator $G1$ creates a realistic vascular mask $M_1$, fitting and generating $M_2$ as a positive sample. Generator $G2$ synthesizes the vascular mask $M_3$ and its synthetic vascular angiography image $I_3$, serving as a negative sample. In addition, XCA means angiographic X-rays. }
  \label{fig:overview}
\end{figure*}
%-------------------------------------------------------------------------
% \subsection{Overview}
Vascular structures have complex geometric properties crucial in blood vessel synthesis. To synthesize reasonable structures of blood vessels, this paper first proposes a parametric vascular model, introduced in Section~\ref{pvm}.  
The network is introduced in Section~\ref{network}, consisting of three modules. Section~\ref{diffusion} introduces the diffusion model for learning vessel feature distributions. Section~\ref{generation} describes the generation module for synthesizing angiographic X-rays. The discriminant module is introduced in Section~\ref{discriminater}. 
At last, Section~\ref{selftraining} presents the self-training pipeline by cyclic consistencies. 
%-------------------------------------------------------------------------
\subsection{Parametric vascular model}\label{pvm}
Blood vessels have a complex geometry.
To better synthesize high-quality vascular angiography with well-observed blood vessels, we propose a parametric vascular model for morphological simulation of the geometric structures. 
Here, we employ adaptive Bézier curves to accurately simulate the branching patterns and geometrical shapes of blood vessels. This model is designed for dual purposes: it constrains the geometry within existing vascular masks, ensuring a precise fit, and facilitates the random generation of vascular masks that adhere to the inherent characteristics of vascular geometry. This approach enhances the fidelity and applicability of vascular simulations in biomedical studies.

\subsubsection{Preliminaries of Bézier curves}\label{bezier}
Ma et al.~\cite{ma2021self} first introduced a vascular parameter shape model, designing a fractal synthesis module based on rectangular blocks to generate various vascular shapes. However, blood vessels generated by this method are not realistic as blood vessels are often curved. 
Thus, it reminds us of using Bézier curves to model such curved structures. 
We test different types of Bézier curves~\cite{bulut2023path,wen2024cad} in Section~\ref{ablations} and adopt cubic Bézier curves. 
It is defined by four pivotal control points $(P_0, P_1, P_2, P3)$, where $P_0$ and $P_3$ are the endpoints of the curve and $P_1$ and $P_2$ dictate its trajectory and form, the cubic Bézier curve is expressed mathematically as:
% \begin{equation}
%     \begin{aligned}
%     \resizebox{.8\hsize}{!}{$
%     B(t)=(1-t)^3 P_0 + 3(1-t)^2 t P_1 + 3(1-t) t^2 P_2 + t^3 P_3
%     $}
%     \end{aligned}
% \end{equation}
\begin{equation}
    \begin{aligned}
    B(t)=(1-t)^3 P_0 + 3(1-t)^2 t P_1 + 3(1-t) t^2 P_2 + t^3 P_3 ,
    \end{aligned}
\end{equation}
where $t$ ranges from 0 to 1. By fitting the parameters of control points, it can intricately simulate the complex geometric structures of blood vessels. Following the characteristics of blood vessels, we encourage one endpoint to lie on another curve, as blood vessels are also branched and connected.

\subsubsection{Synthetic blood vessel genration}\label{vascularfitting0}
This section introduces the synthesis of blood vessel masks using a parametric vascular model.

\textbf{Initialisation of the vascular skeleton.}
Firstly, we synthesize a vascular skeleton to guide the growth of Bézier curves. Initializing the vascular scaffold, we employ a tree structure to represent the vascular system. Initially, we define a root node, symbolizing the starting point of the entire vascular system, which encompasses basic information about the vessel. Subsequently, we construct the branching architecture of vessels. Each node signifies a vascular segment, containing characteristics of the segment such as length and connectivity. The vascular system is decomposed into increasingly smaller vessel segments through a progressive bifurcation process. We establish a maximum depth parameter to determine the maximum branching number, representing the peripheral vessels' reach. In this process, a recursive algorithm is utilized to build the tree-like structure of the entire vascular system. Ultimately, we achieve a comprehensive vascular scaffold structure where the root node represents the largest vessel, the leaf nodes denote the terminal ends of vessels, and intermediate nodes symbolize the various levels of vascular branches.

\textbf{Adaptive Bézier curve generation.}
To enhance the initialization of the vascular mask, our methodology incorporates adaptive Bézier curve modeling. Adjusting to the local attributes of the vascular structure offers more precise simulations than conventional methods that depend on static parameters to generate curves. For each node position along the vascular skeleton, a method predicated on the characteristics of the surrounding area is utilized to ascertain the Bézier curve's direction. In particular, the gradient information regarding length and depth for different branches at a node position is determined via the Sobel operator, facilitating the calculation of the angle characteristic of the local region. Bézier curves, formulated based on this angle, ensure proper alignment with the vascular framework.

\begin{algorithm}
\caption{Vascular mask generation by the parametric vascular model}\label{alg:vessel}
\begin{algorithmic}[1]
\Require Vascular skeleton $\mathcal{S}$, depth parameter $\omega$, number of branches $\theta$, tree nodes $x_0, x_1, \ldots, x_n$
\Ensure Initialised geometry of a blood vessel mask $F_i$
\State Initialize $\xi \gets 0$
\While{$i \leq \omega$} 
    \Repeat
        \State Randomly initialize node positions and lengths
        \While{$\xi \neq 0$}
            \Repeat
                \State \textbf{Step 1:} Perform morphological erosion $\mathcal{T} \gets \text{Erosion}(\mathcal{S})$
                \State \textbf{Step 2:} Perform morphological dilation $\mathcal{D} \gets \text{Dilation}(\mathcal{T})$
                \State \textbf{Step 3:} Compute bitwise OR operation $\mathcal{R} \gets \mathcal{T} \lor \mathcal{D}$
            \Until{$\mathcal{T} = 1$}
            \State Update $\xi \gets 0$
        \EndWhile
    \Until{$x_0, x_1, \ldots, x_n \in \mathcal{S}$}
\EndWhile
\State Initialize Bézier curve $\psi$ with four control points $\psi_1, \psi_2, \psi_3, \psi_4$
\State Initialize skeleton tree points $\rho_1, \rho_2, \ldots, \rho_n$
\While{$F_i$ \textbf{does not} cover $\mathcal{S}$}
    \While{$\psi \notin \mathcal{S}$}
        \State Initialize Bézier curve, ensuring $\psi_1, \psi_4 \in \mathcal{S}$
        \Repeat
            \State Compute Sobel gradient $\frac{\partial \rho_i}{\partial \psi}$
            \State Optimize $\psi_2, \psi_3$ such that $\theta = \sqrt{\rho_n, \psi}$ minimizes $\theta$
            \State Update Bézier curve size by $\omega$
        \Until{$\min(\psi) \in \alpha$}
    \EndWhile
    \State Update $\psi_4 \gets$ last $\psi_1$ for the next Bézier curve
\EndWhile
\end{algorithmic}
\end{algorithm}

%-------------------------------------------------------------------------
% \vspace{-0.5cm}
% \subsection{Our network}\label{network}
\subsection{Self-supervised adversarial diffusion model \label{network}}
This section introduces the networks and the self-supervised training scheme. The overall pipeline can synthesize vascular angiography from non-angiographic X-rays by training on unpaired and unlabeled data. 

The network consists of three modules. 
Firstly, a diffusion model is adopted to model the data distribution and transformation from non-angiographic X-rays to angiographic X-rays, learning from unpaired data. Details such as blood vessels are well observed in angiographic X-rays, but they may be invisible in non-angiographic X-ray images. Thus, blood vessels can be further generated, guided by synthesized vascular masks. 
The generation module synthesizes realistic blood vessels on generated vascular angiography. 
Lastly, an adversarial discriminant enforces the quality of generated vascular angiography. It consists of a mask discriminator and an image discriminator. 
The mask discriminator distinguishes between synthesized vascular masks and real vascular masks, while the image discriminator distinguishes between synthesized vascular images and real vascular images. This structure promotes vessel feature learning and enhances the capability of the generation module. 

The overview of this method is presented in Figure \ref{fig:overview}. 
Firstly, we input unpaired vascular angiography $I_1$ and non-angiographic X-ray $I_2$ into a denoised diffusion probability model (DDPM) to learn the characteristic distribution of blood vessels. The noisy data distribution map is used as input to the generator module. Meanwhile, the vascular morphogenetic module is utilized to randomly initialize the initial vascular mask of the vascular geometric features as the attention input of the generator. The generator will output $mask_1$ of $I_1$ and the resultant vascular image $I_3$, along with the corresponding mask $mask_3$. By fitting the parametric vascular model, we use $mask_1$ to obtain a model-fitted  $mask_2$, and feed to the discriminator as positive samples.
$Mask_3$ is served as a negative sample for the discriminator $D1$. 
At the same time, $I_1$ is served as a positive sample, and the synthetic vascular angiography $I_3$ is served as a negative sample. These inputs are fed into the discriminator $D2$ for authenticity judgments.

\subsubsection{Diffusion model}\label{diffusion}
To explore the distribution of blood vessels, we utilize unpaired vascular angiography $I_1$ and non-angiographic X-ray $I_2$ as inputs for the diffusion model. While the forward passes, the module gradually adds noises to simulate the transformation from vascular angiography with clear blood vessels into a noisy state through a parameterized Markov process. This process is shown in Eq. \eqref{eq:eq_markov}. 
\begin{equation}\label{eq:eq_markov}
x_t=\sqrt{1-\beta_t} x_{t-1}+\sqrt{\beta_t} \epsilon,
\end{equation}
where $\epsilon$ corresponds to noise extracted from a standard normal distribution, while $\beta_t$ signifies a pre-defined noise magnitude parameter, which regulates the quantity of noise superimposed at each procedural iteration. From $t=1$ to $t=T$ ($T$ is the end time of the diffusion process), Loop through the above steps until you get a fully noisy image $x_t$.

The reverse denoising process is a crucial training phase, enabling the extraction of vital features from vascular images through denoising. This process involves starting with $x_t$ and defining a denoising function $f_\theta$, which predicts the denoised image $x_{t-1}$ based on the current noisy state $x_t$ and the time step to $t$. The process is shown in Eq. \eqref{eq:eq_markov2} and Eq. \eqref{eq:eq_markov3}.
\begin{equation}\label{eq:eq_markov2}
x_{t-1}=f_\theta\left(x_t, t\right),
\end{equation}
\begin{equation}\label{eq:eq_markov3}
% x_{t-1}=x_t-\sigma \nabla f_\theta\left(x_t, t\right),
x_{t-1} = f_\theta(x_t, t) + \sigma_t \epsilon_t,
\end{equation}
where $f_\theta(x_t, t)$ is the predicted denoised image at step $t$, and $\sigma_t \epsilon_t$ adjusts the noise, with $\sigma_t$ controlling its scale and $\epsilon_t$ being the model's noise estimate.

The diffusion loss is utilized during the training of the diffusion model to enable effective learning of the data distribution of the vasculature. It promotes feature consistency and enhances the model's capacity to capture complex patterns. For a given vascular image denoted as $x_0$, along with noise represented by $\epsilon$, $\epsilon_\theta$ is the noise predicted by the diffusion model, and at a specific time step $t$, the diffusion loss is computed by comparing the model's output at time step $t$ with the actual noise. is precisely defined in Eq. \eqref{eq:diff_loss}.
\begin{equation}\label{eq:diff_loss}
L_{diff} = \mathbb{E}_t \left[ \left\| \epsilon - \epsilon_\theta \left( \sqrt{\alpha_t} x_0 + \sqrt{1 - \alpha_t} \epsilon, t \right) \right\|^2 \right].
\end{equation}

\subsubsection{Generation module}\label{generation}
\begin{figure}[ht]
  \centering
  \includegraphics[width=\linewidth]{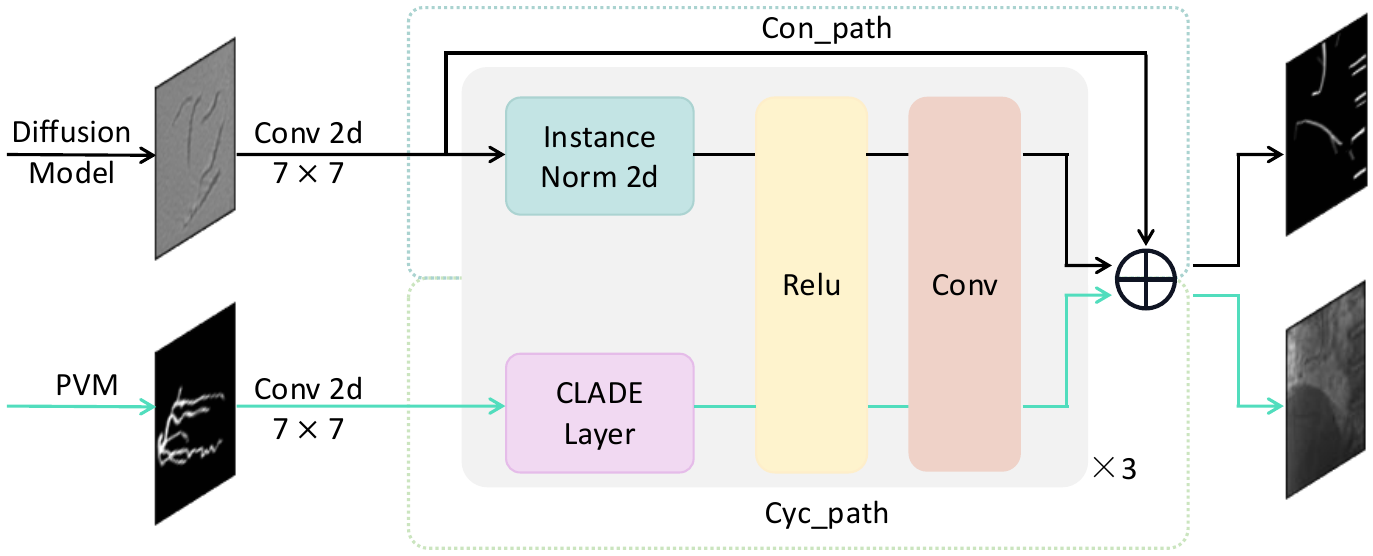}
  \caption{Generator generates vessel masks via the con\_path when the vessel mask is unavailable or synthesizes vessel images via the cyc\_path when the mask is available.}
  % \vspace{-0.3cm}
  \label{fig:generate}
\end{figure}

The generator module actively receives the noise data distribution map from the diffusion model and the initial vascular mask generated by the proposed vascular model. It then synthesizes both the angiography and vascular masks based on these inputs. Specifically, the generator module consists of N residual blocks, which utilize configurable class-adaptive normalization (CLADE) layers~\cite{tan2021efficient} to generate either vascular masks or synthesized vascular angiography selectively. As shown in Figure \ref{fig:generate}, the process in the generator module is divided into two paths: the mask synthesis path (Con\_path in Figure \ref{fig:generate}.) and the image synthesis path (Cyc\_path in Figure \ref{fig:generate}.).

The network architectures of \textit{G1} and \textit{G2} are identical. In the absence of an initial vascular mask, \textit{G1} executes the contrast path (Con\_path), integrates the Parametric Vascular Model (PVM) and applies instance normalization (Instance Norm) to generate an initial vascular mask that initializes the contrast path.

During the formal angiography synthesis process, \textit{G2} activates the instance normalization (Instance Norm) in the contrast path (Con\_path) to generate a new angiography mask. Simultaneously, the CLADE layer in the adversarial path (Cyc\_path) is activated to produce a synthetic image. The vascular mask is then used as semantic input to guide the optional CLADE layer in generating a semantic angiography image. Additionally, using CLADE to synthesize vascular representations in the adversarial path aids in estimating the geometric structure of the angiography within the contrast path.

Mask-based contrastive loss and cycle loss are employed in the generator, where the mask-based contrastive loss ensures the similarity between the actual vascular $mask_2$  and the synthesized mask $mask_3$. Cycle loss is modeled optimally within the Cyc\_path, ensuring accurate angiography is synthesized. The Mask-based contrastive loss is specified in Eq. \eqref{eq:mask_loss}, and the cycle loss is specified in Eq. \eqref{eq:cycle_loss}.
\begin{equation}\label{eq:mask_loss}
% \begin{aligned}
% \resizebox{.95\hsize}{!}{$
{L}_{mask}= \mathbb{E}_{m, \hat{m}_v, \hat{m}_f}\left[\sum_{r=1}^R \sum_{q=1}^{Q_r} \ell_{\mathrm{MI}}\left(h_q^r\left(m_i\right), h_q^r\left(\hat{m}_i\right), h_q^r\left(\hat{m}_v\right)\right)\right] ,
% }
% \end{aligned}
\end{equation}
where ${h}_{q}^{r}$ represents the noisy data distribution. $m$ denotes the initial vascular mask, while $\hat{m}_v$ and $\hat{m}_i$ respectively denote the synthesized vascular mask and the initial vascular mask. $\ell_{\mathrm{MI}}$ signifies the mutual information computed via cross-entropy.
\begin{equation}\label{eq:cycle_loss}
{L}_{cyc}= \mathbb{E}_{x_t, m}\left[\left\|G\left(G\left(x_t, m\right), 0\right)-m\right\|_1\right],
\end{equation}
where \textit{G} represents the output of the generator, and $m$ denotes the synthetic mask.

\subsubsection{Adversarial discriminant module}\label{discriminater}
We have two discriminators in the adversarial discriminant module. The first is the mask discriminator ($D1$), which aims to differentiate between model-generated vascular masks and real vascular masks. 
By taking generated masks and real vascular masks as negative and positive samples, $D1$ endeavors to ascertain their authenticity. 
The second one is the angiographic X-ray discriminator ($D2$), tasked with distinguishing synthetic angiographic X-rays generated by the model from genuine angiographic X-rays. %$D2$ assesses the credibility of the resulting composite image.

By this framework, we employ a multi-layer CNN known as NLayerDiscriminator as the discriminator. This network is specifically designed to efficiently extract and process image features, supporting adversarial learning tasks. 

Adversarial loss is computed by minimizing the disparity between images generated by the generator and real images. It can be represented in Eq. \eqref{eq:adv_loss}:
\begin{equation}\label{eq:adv_loss}
\begin{split}
{L}_{adv}^G & = \mathbb{E}_{x_t, m_i}\left[\left(D\left(G\left(x_t, m_i\right)\right)-1\right)^2\right], \\
{L}_{adv}^D & = \frac{1}{2} \mathbb{E}_{x_0}\left[\left(D\left(x_0\right)-1\right)^2\right] 
+ \frac{1}{2} \mathbb{E}_{x_t, m_i}\left[\left(D\left(G\left(x_t, m_i\right)\right)\right)^2\right], \\
{L}_{adv} & = {L}_{adv}^G + {L}_{adv}^D ,
\end{split}
\end{equation}
where \textit{D} is the output of the discriminator,$x_t$ is noisy data distribution maps, and $m_i$ is initialized mask.

\subsubsection{Self-training for cyclic consistency}\label{selftraining}
We further adopt cyclic consistencies to improve the performance in a self-supervised manner. Firstly, the diffusion model processes non-angiographic X-rays to obtain a noisy data distribution map. This map is then fed into the generator module. Guided by the vascular mask, the system synthesizes both the angiographic X-rays $I_3$ and the corresponding vascular $mask_3$, which serve as negative samples for the adversarial network. 
The parametric vascular model is also applied to re-fit the fake mask. Subsequently, fake angiographic X-rays and fake masks are reintroduced to the diffusion model for self-training. 
This iterative process ensures that the generated angiographic X-rays are consistent with the original non-angiographic image.

The model introduces a cycle consistency loss by calculating the difference between the original vascular mask and the reconstructed synthetic mask, ensuring the generated vascular mask accurately reflects the original vascular structure. 
By minimizing this loss, the model will more precisely maintain the details of the vascular structure without an explicit supervisory signal, improving the synthesis quality. Cyclical consistency not only serves as a regularization technique but also enhances the understanding of image features, thus improving the generalization capability and robustness in vascular synthesis tasks. 
Therefore, the total loss function of the model can be defined as shown in Eq. \eqref{eq:total_loss}.
\begin{equation}\label{eq:total_loss}
    \textit{L}={L}_{diff}(\epsilon)+\lambda_\alpha{{L}_{mask}}+\lambda_\beta {L}_{adv}+\lambda_\gamma {L}_{cyc},
\end{equation}
where $\lambda_\alpha$, $\lambda_\beta$, $\lambda_\gamma$ are hyper-parameters.

\section{Experiments}
%-------------------------------------------------------------------------
\subsection{Dataset}
\textbf{XCAD dataset.}
It is captured using the General Electric Innova IGS 520 system at a resolution of 512x512~\cite{ma2021self}. We divided the data into three categories (training set, validation set, and test set), and ensured that there was no sample overlap between different data sets. The training and validation set comprises 1621 angiographic X-rays without mask annotations and unpaired non-angiographic X-rays. We use these non-angiographic X-rays as $I_2$ in Figure~\ref{fig:overview}. The test set comprises 126 independent angiographic X-rays along with masks, where professional radiologists have annotated the vascular segmentation maps. We use the test set to evaluate the fitting performance of the proposed parametric vascular model in Section~\ref{vascularfitting}. 

\textbf{ARCADE dataset.}
It is proposed by Popov et al.~\cite{popov2024dataset} and improves the diagnosis of coronary artery disease using region-based approaches. It uses angiographic X-rays to segment coronary arteries, detect stenoses, and diagnosis.  The stenosis portion of this dataset includes 1500 annotated angiographic X-rays. They highlight coronary artery segments and stenotic plaque locations. Similarly, the syntax section contains 1500 annotated images with different coronary artery segments. We augment angiographic X-rays from ARCADE and XCAD to serve as $I_1$ in Figure~\ref{fig:overview}. 

\textbf{SynthXCA dataset.}
Our method synthesizes corresponding angiographic X-rays from non-angiographic X-rays. The generated angiographic X-rays ($I_3$ in Figure~\ref{fig:overview}) and corresponding vascular masks ($M_3$ in Figure~\ref{fig:overview}), along with paired non-angiographic X-ray images $I_2$ compose the SynthXCA dataset. There are 1500 data in total. 

%-------------------------------------------------------------------------
\subsection{Comparison of Parameterized Vascular Models}\label{vascularfitting}
In Section~\ref{pvm}, we propose a parametric vascular model to fit the vascular geometry. To validate its effectiveness, we perform vascular fitting compared with a state-of-the-art vascular model~\cite{ma2021self}. We evaluate the quality of vascular fitting by comparing it with annotated vessels using two datasets. The first dataset contains 126 groups of real data extracted from the test part of the XCAD dataset, consisting of labeled vessel masks for fitting. The second dataset comprises 151 sets of synthetic data derived from the labeled raw dataset presented by Qin et al. in RPCA-UNet~\cite{qin2022robust}. 
IOU~\cite{yu2016unitbox}, SSIM~\cite{wang2004image}, and MSE are used as metrics to assess the similarity between fitted and labeled vessel masks.

We compare our method with the fractal vascular model by Ma et al.~\cite{ma2021self}. They designed the model to simulate vascular masks as well. We compare both models by fitting them to annotated masks and comparing their fitting performance. 

\begin{figure*}[t]
  \centering
  \includegraphics[width=\linewidth]{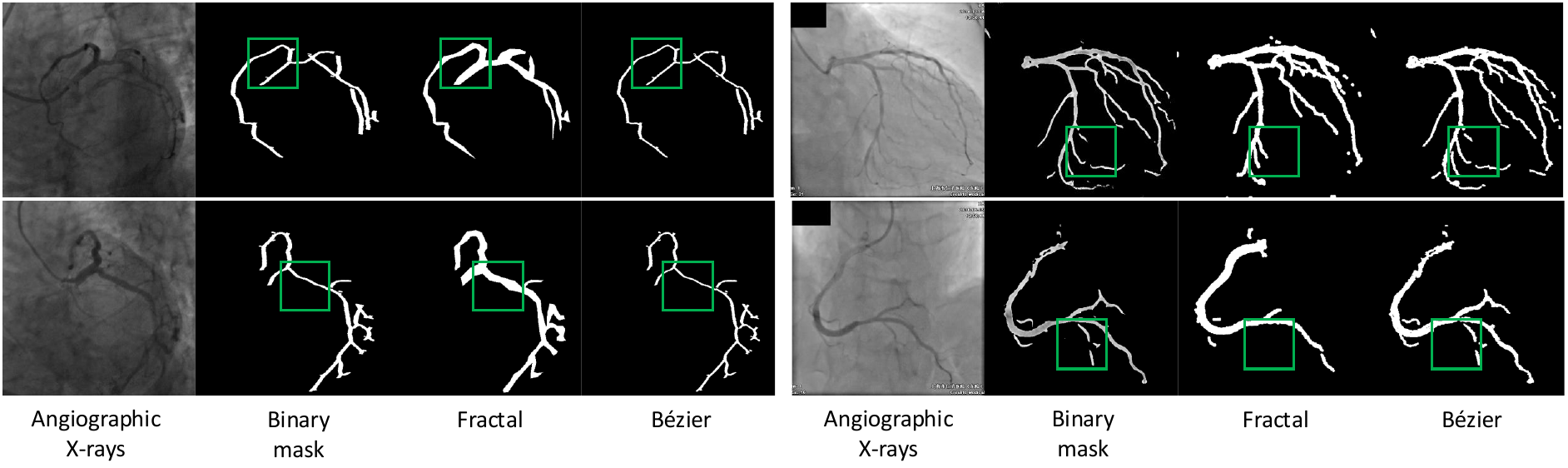}
  \caption{Fitting blood vessel masks via Fractal and Cubic Bézier curve method. XCADeval (left) and VesMask-RPCA (right).}
  \label{fig:bezier_v1}
\end{figure*}

In Table \ref{tab:pvm-comparisontab}, we can observe that our parametric vascular model consistently outperforms the fractal model, achieving better performance. In Figure \ref{fig:bezier_v1}, we show the visual comparisons of vascular fitting. In the green boxed area, fractals exhibit an unnatural rectangular effect in their geometric appearance, 
missing segments in certain small blood vessels. Fitting by our curve model is closer to annotated masks.
% \begin{table}
% \centering
% \caption{Metrics for Fractal and PVM vascular mask fitting.}
% \label{tab:pvm-comparisontab}
% \small
% \setlength{\tabcolsep}{6pt} % Adjust column spacing to fill the single column width
% \resizebox{\linewidth}{!}{
% \begin{tabular}{c|ccc|ccc}
% \hline
%  & \multicolumn{3}{c|}{XCADeval} & \multicolumn{3}{c}{VesMask-RPCA} \\ \cline{2-7} 
%  & IOU $\uparrow$ & SSIM $\uparrow$ & MSE $\downarrow$ & IOU $\uparrow$ & SSIM $\uparrow$ & MSE $\downarrow$ \\ \hline
% Fractal~\cite{ma2021self} & 0.600 & 0.893 & 0.037 & 0.636 & 0.888 & 0.079 \\ \hline
% PVM & \textbf{0.617} & \textbf{0.920} & \textbf{0.025} & \textbf{0.669} & \textbf{0.927} & \textbf{0.076} \\ \hline
% \end{tabular}
% }
% \end{table}

\begin{table}[ht]
\caption{Metrics for Fractal and PVM vascular mask fitting.}\label{tab:pvm-comparisontab}
\begin{tabular*}{\textwidth}{@{\extracolsep\fill}lccc|ccc}
\toprule
\multirow{2}{*}{Methods} & \multicolumn{3}{c|}{XCADeval} & \multicolumn{3}{c}{VesMask-RPCA} \\
\cmidrule{2-7}
                        & IOU $\uparrow$  & SSIM $\uparrow$ & MSE $\downarrow$ & IOU $\uparrow$  & SSIM $\uparrow$ & MSE $\downarrow$ \\
\midrule
Fractal~\cite{ma2021self} & 0.600 & 0.893 & 0.037 & 0.636 & 0.888 & 0.079 \\
PVM & \textbf{0.617} & \textbf{0.920} & \textbf{0.025} & \textbf{0.669} & \textbf{0.927} & \textbf{0.076} \\
\botrule
\end{tabular*}
\footnotetext{Note: IOU, SSIM, and MSE are used as metrics to evaluate the performance of vascular mask fitting. $\uparrow$ indicates higher is better, and $\downarrow$ indicates lower is better.}
\end{table}

%-------------------------------------------------------------------------
\subsection{Method evaluations}
To evaluate the image quality of angiography synthesis, this paper proposes a comparison with methods based on modality conversion. Additionally, to objectively evaluate the geometric quality of synthesized angiography, this work analyzed the clinical angiography process. Due to factors such as imaging technology and uneven contrast agent injection, angiography may exhibit geometric discontinuities or breaks. Additionally, issues such as noise, motion artefacts, or insufficient imaging resolution can lead to unsmooth or blurred vessel boundaries. These are critical challenges in angiographic imaging. To address these, we designed evaluation metrics from the perspectives of vascular connectivity and edge smoothness, as detailed in Section~\ref{metrics}.

\subsubsection{Metrics}\label{metrics}
In terms of image synthesis quality, we perform a comparative evaluation using PSNR and SSIM. For vascular geometric structure, we design metrics for vascular connectivity and edge smoothness.

% \paragraph{Vascular Connectivity Metrics}
Vascular connectivity metrics assess the integrity and continuity of vascular structures through skeletonization. The canny edges algorithm is employed to extract vessels, simplifying the vessel shape while preserving its topological features. To quantify vascular connectivity, we propose Connectivity Ratio (\textit{CR}) metrics.
\begin{equation}
\textit{CR} = \frac{\textit{sk-pixels}}{\textit{vas-pixels}} \times 100\%,
\end{equation}
where $sk-pixels$ stand for connected skeleton pixels, $vas-pixels$ stand for total vascular pixels, and $CR$ is better when larger, as it indicates stronger connectivity. 

% \paragraph{Vascular Edge Smoothness Metrics}
Vascular edge smoothness metrics evaluate the smoothness of vessel boundaries. Edge smoothness is measured by the curvature of the vascular boundary, defined as the second derivative:
\begin{equation}
S_{\text{smooth}} = \frac{d^2y}{dx^2},
\end{equation}
where the curvature ($S_{\text{smooth}}$), a smaller value indicates a smoother, less curved boundary. 

\subsubsection{Evaluation results}
\begin{figure*}[t]
  \centering
  \includegraphics[width=\linewidth]{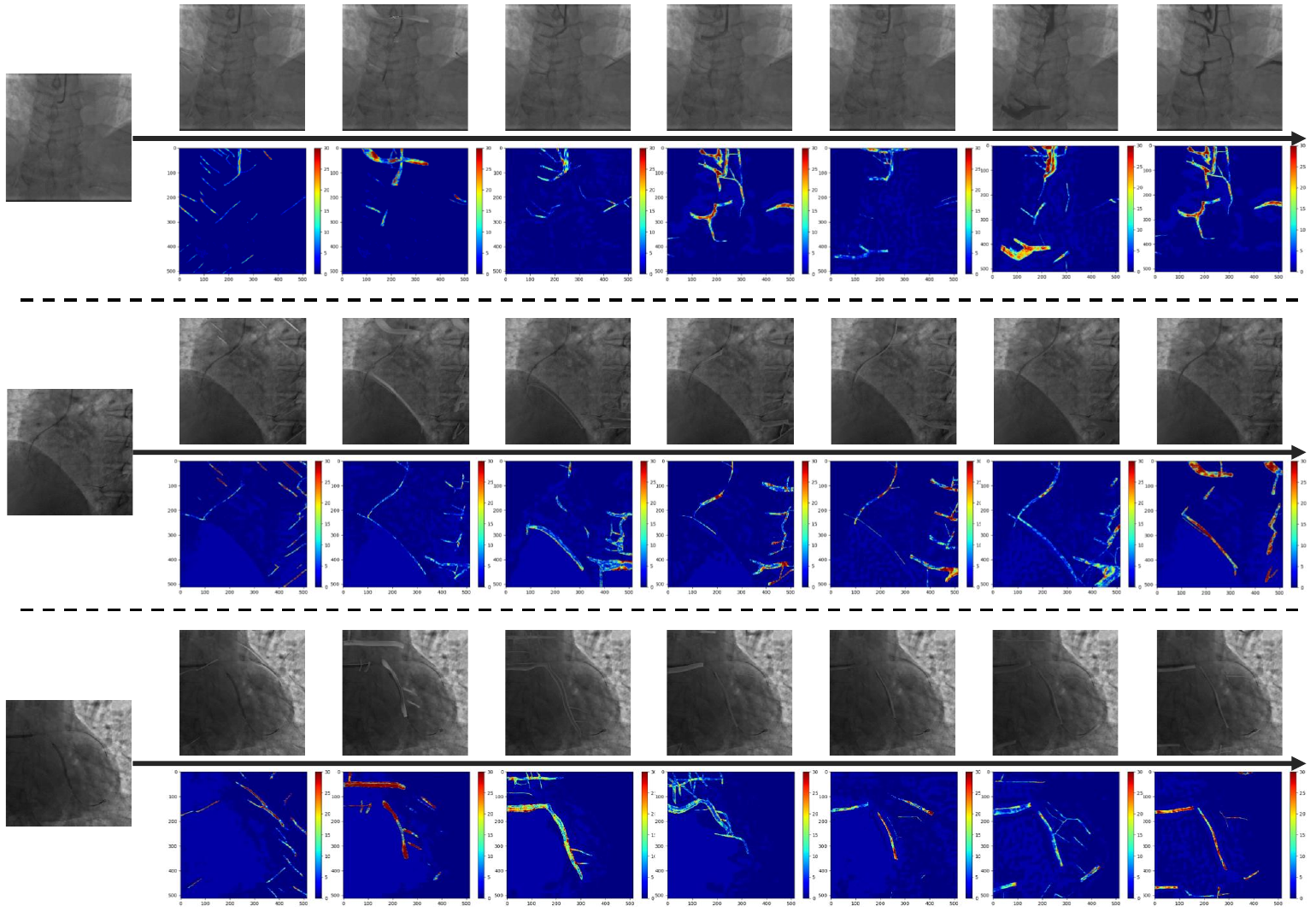}
  \caption{Examples of the process of synthesizing angiographic X-rays. For each set of examples, the first row shows the updated angiographic X-rays. The second row of heat maps highlights the differences between the synthesized angiographic images and the non-angiographic X-rays.}
  \label{fig:bezier_v2}
  \vspace{-0.3cm}
\end{figure*}

Figure \ref{fig:bezier_v2} shows the angiographic X-ray and thermogram results. In the first row of X-rays, some areas are prone to misclassification. However, with more adversarial training iterations, non-vascular regions are excluded, allowing the network to focus on realistic vascular synthesis. Additionally, the vessel geometry becomes smoother and more natural as training progresses.

\begin{figure}
  \centering
  \includegraphics[width=\linewidth]{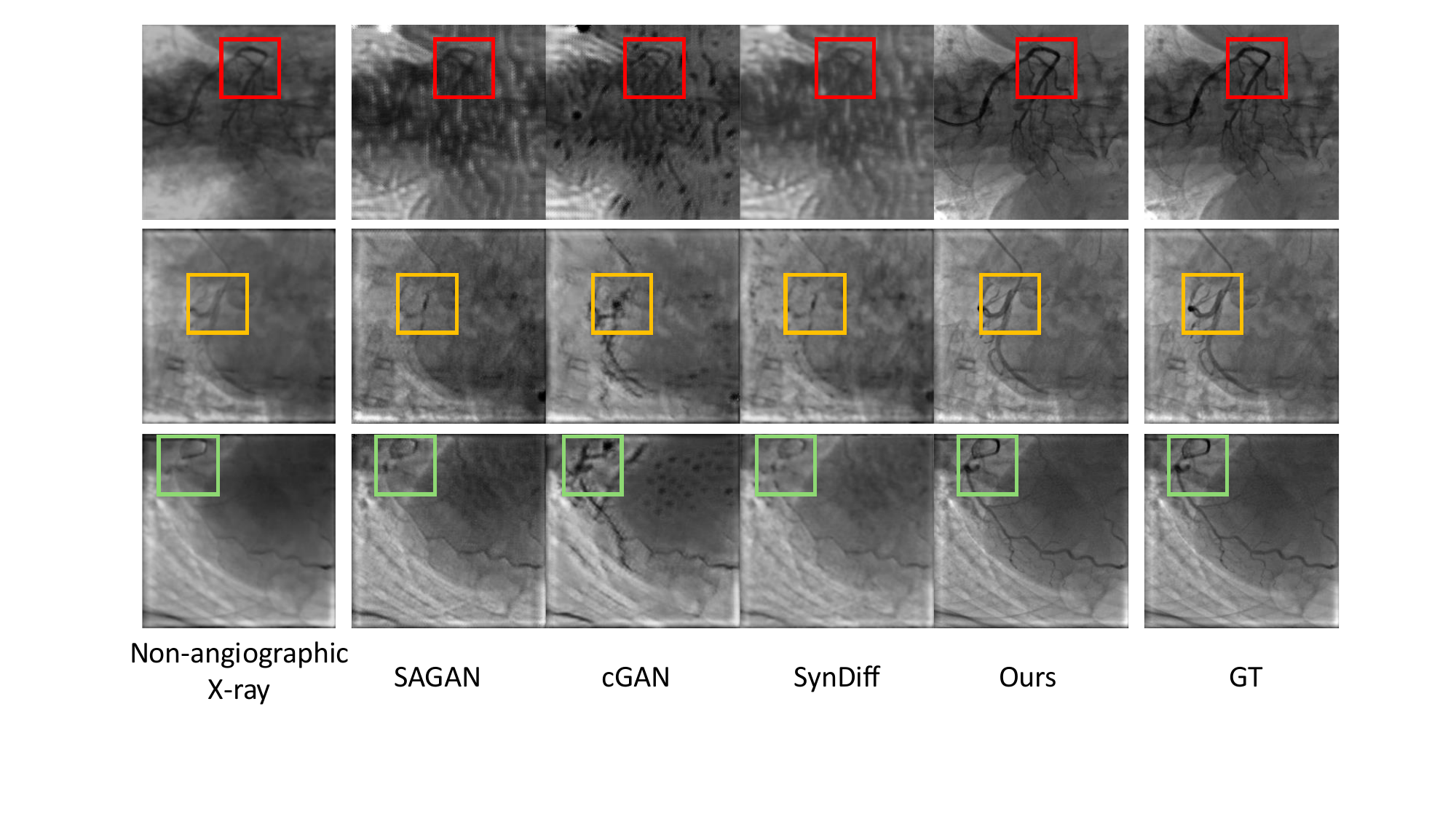}
  \caption{Comparison of angiographic vessels from non-angiographic vessels with other methods.}
  \label{fig:result_cross}
  \vspace{-0.3cm}
\end{figure}

To evaluate the synthesis quality of our method in angiography, particularly for vascular geometry, we treated angiography synthesis as a modality conversion task from vascular X-ray to XCA due to the lack of dedicated methods. To assess this, we compared our method with SAGAN~\cite{zhang2019self}, cGAN~\cite{dar2019image}, and Syndiff~\cite{ozbey2023unsupervised} using the train-B and train-C subsets of the partially processed XCAD dataset. The data preprocessing pipelines of methods such as Syndiff were modified to better adapt to the characteristics of vascular X-rays and angiographic X-ray images. The synthesis performance was evaluated using PSNR and SSIM metrics, with the results presented in Table \ref{tab:cross} and Figure \ref{fig:result_cross}. Our experiments demonstrate that, while achieving comparable angiography quality, our method generates vascular geometries that are more continuous and smoother.

To better meet clinical needs, we evaluated the model parameters and inference time of various methods, as shown in Table \ref{tab:times}. The results demonstrate that our method achieves faster inference and requires fewer parameters, indicating its strong potential for clinical applications. This is attributed to the fact that our method leverages the forward diffusion process to learn the latent features of angiography and X-ray images and utilizes an optional generator to produce angiograms, without relying on the reverse diffusion process for direct generation.

% \begin{table}
% \centering
% \caption{Comparison of synthesis results with other medical modality translation methods.}
% \label{tab:cross}
% \small
% \resizebox{\linewidth}{!}{
% \begin{tabular}{c|c|c|c|c}
% \hline
% Methods & SAGAN~\cite{zhang2019self}  & cGAN~\cite{dar2019image}   & SynDiff~\cite{ozbey2023unsupervised} & Ours   \\ \hline
% PSNR $\uparrow$    & 20.63  & 24.78  & 25.46   & \textbf{26.42}  \\ \hline
% SSIM $\uparrow$   & 0.6547 & 0.8163 & 0.8532  & \textbf{0.8648} \\ \hline
% $CR$ $\uparrow$      & 0.6012 & 0.6281 & 0.7218  & \textbf{0.8213} \\ \hline
% S_{\text{smooth}} $\downarrow$       & 0.4823 & 0.4513 & 0.4217  & \textbf{0.3511} \\ \hline
% \end{tabular}}
% \end{table}
\begin{table}[ht]
\caption{Comparison with other modality translation methods on XCAD dataset subset.}\label{tab:cross}
\begin{tabular*}{\textwidth}{@{\extracolsep\fill}lcccc}
\toprule%
Methods & SAGAN & cGAN & SynDiff & Ours \\
\midrule
PSNR $\uparrow$    & 20.63  & 24.78  & 25.46   & \textbf{26.42}  \\
SSIM $\uparrow$    & 0.6547 & 0.8163 & 0.8532  & \textbf{0.8648} \\
$CR$ $\uparrow$    & 0.6012 & 0.6281 & 0.7218  & \textbf{0.8213} \\
$S_{\text{smooth}}$ $\downarrow$ & 0.4823 & 0.4513 & 0.4217  & \textbf{0.3511} \\
\botrule
\end{tabular*}
\end{table}

% \begin{table}[]
% \centering
% \caption{Comparison of parameters and Inf. Time across methods.}
% \label{tab:times}
% \small
% \setlength{\tabcolsep}{2pt} % Adjust column spacing to fill the single column width
% \resizebox{\linewidth}{!}{
% \begin{tabular}{c|c|c|c|c}
% \hline
% Method        & SAGAN~\cite{zhang2019self} & cGAN~\cite{dar2019image}  & SynDiff~\cite{ozbey2023unsupervised} & Ours  \\ \hline
% Parameters $(M) \downarrow$ & 58.32 & 52.66 & 47.54   & \textbf{35.48} \\ \hline
% Time $(s) \downarrow$       & 0.415 & 0.324 & 0.126   & \textbf{0.102} \\ \hline
% \end{tabular}}
% \vspace{-0.5cm}
% \end{table}
\begin{table}[ht]
\caption{Comparison of parameters and inference time across methods.}\label{tab:times}
\begin{tabular*}{\textwidth}{@{\extracolsep\fill}lcccc}
\toprule
Methods & SAGAN & cGAN & SynDiff & Ours \\
\midrule
Parameters $(M) \downarrow$ & 58.32 & 52.66 & 47.54 & \textbf{35.48} \\
Time $(s) \downarrow$       & 0.415 & 0.324 & 0.126 & \textbf{0.102} \\
\botrule
\end{tabular*}
\end{table}

%-------------------------------------------------------------------------
\subsection{Ablation studies}\label{ablations}
\subsubsection{Different order of Bézier curves}\label{diff_bezier}
We propose a parametric vascular model employing a cubic Bézier curve (CB) to simulate the geometry of blood vessels. To assess the impact of Bézier curve order on the fitting quality of vascular structures, we are conducting an ablation study, testing three configurations: cubic Bézier curves generated via four control points, quartic Bézier curves (QB) defined by five control points, and quintic Bézier curves (QB5) using six control points. 

\begin{figure*}
  \centering
  \includegraphics[width=\linewidth]{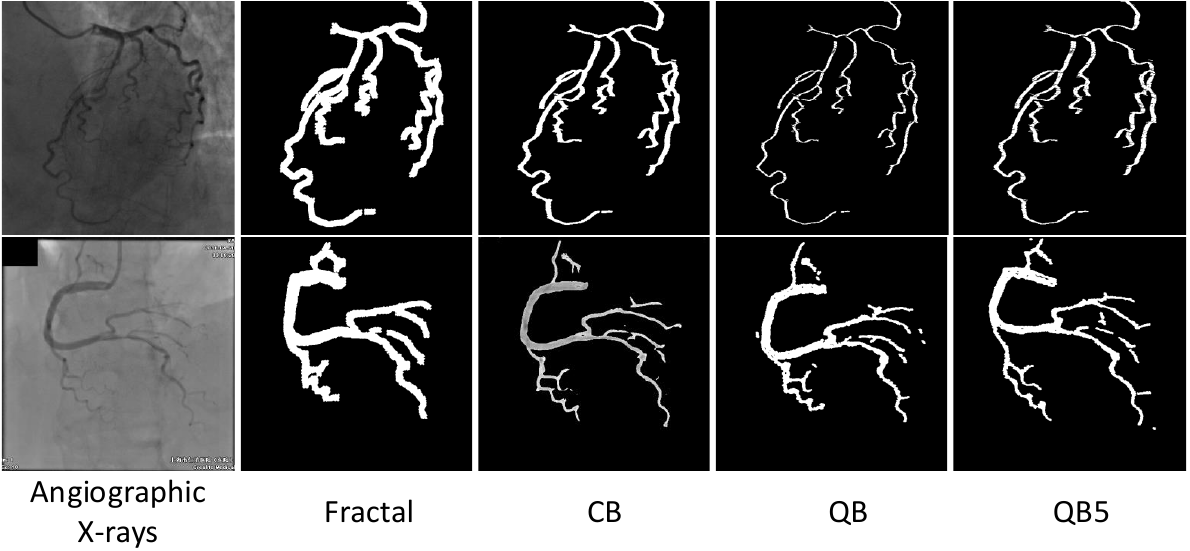}
  \caption{Comparison of fitting effects: original angiographic X-rays, fractal, cubic, quartic, and quintic curves.}
  \label{fig:vmm_eval2}
  \vspace{-0.3cm}
\end{figure*}

As shown in Table \ref{tab:vmm_eval2}, we compare the fitting performance of these curves on the XCADeval and VesMask-RPCA datasets. The results indicate that the cubic Bézier curve yields superior fitting metrics, demonstrating its effectiveness in vascular modeling. The effect is shown in Figure \ref{fig:vmm_eval2}.
% \begin{table}
% \centering
% \caption{Ablations of different orders of Bézier curves.}
% \label{tab:vmm_eval2}
% % \resizebox{\linewidth}{!}{
% \begin{tabular}{c|ccc|ccc}
% \hline
% \multirow{2}{*}{} & \multicolumn{3}{c|}{XCADeval} & \multicolumn{3}{c}{VesMask-RPCA} \\ \cline{2-7} 
%                   & \multicolumn{1}{c|}{IOU} & \multicolumn{1}{c|}{SSIM} & MSE & \multicolumn{1}{c|}{IOU} & \multicolumn{1}{c|}{SSIM} & MSE \\ \hline
% CB                & \multicolumn{1}{c|}{\textbf{0.617}} & \multicolumn{1}{c|}{\textbf{0.920}} & \textbf{0.025} & \multicolumn{1}{c|}{\textbf{0.769}} & \multicolumn{1}{c|}{\textbf{0.927}} & \textbf{0.076} \\ \hline
% QB                & \multicolumn{1}{c|}{0.407} & \multicolumn{1}{c|}{0.916} & 0.034 & \multicolumn{1}{c|}{0.692} & \multicolumn{1}{c|}{0.914} & 0.083 \\ \hline
% QB5               & \multicolumn{1}{c|}{0.527} & \multicolumn{1}{c|}{0.918} & 0.029 & \multicolumn{1}{c|}{0.677} & \multicolumn{1}{c|}{0.904} & 0.091 \\ \hline
% \end{tabular}
% \vspace{-0.3cm}
% \end{table}
\begin{table}[ht]
\caption{Ablation study of different orders of Bézier curves.}\label{tab:vmm_eval2}
\begin{tabular*}{\textwidth}{@{\extracolsep\fill}lccc|ccc}
\toprule
\multirow{2}{*}{Methods} & \multicolumn{3}{c|}{XCADeval} & \multicolumn{3}{c}{VesMask-RPCA} \\
\cmidrule{2-7}
                        & IOU $\uparrow$  & SSIM $\uparrow$ & MSE $\downarrow$ & IOU $\uparrow$  & SSIM $\uparrow$ & MSE $\downarrow$ \\
\midrule
CB                     & \textbf{0.617}  & \textbf{0.920}  & \textbf{0.025}   & \textbf{0.769}  & \textbf{0.927}  & \textbf{0.076}   \\
QB                     & 0.407           & 0.916           & 0.034            & 0.692           & 0.914           & 0.083            \\
QB5                    & 0.527           & 0.918           & 0.029            & 0.677           & 0.904           & 0.091            \\
\botrule
\end{tabular*}
\end{table}
\subsubsection{Loss functions} 
In this paper, we propose minimizing the losses in Eq. \eqref{eq:total_loss} to train our model. To better verify the validity of each loss, we conduct ablation experiments on individual losses. 
We perform ablations for diffusion loss in Table~\ref{tab:diff_xiaorong}. 
It is confirmed that noisy data distribution maps are more effective in extracting the semantic features of blood vessels when diffusion loss is employed compared to when this loss is not used. Mask-based contrast loss and circulation loss are crucial in generating modules for reconstructing vascular angiography with accurate geometric structures. Additionally, adversarial loss significantly enhances the realism of the synthesized angiographic X-rays.
% \begin{table}[]
% \centering
% % \vspace{-0.3cm}
% \caption{Comparative ablation results of different loss functions.}
% \label{tab:diff_xiaorong}
% \begin{tabular}{c|c|c|c|c}
% \hline
%           & PSNR $\uparrow$  & SSIM $\uparrow$   & $CR$ $\uparrow$     & $S_{\text{smooth}}$ $\downarrow$      \\ \hline
% w/o \({L}_{\textit{diff}}\) & 23.65 & 0.7642 & 0.6942 & 0.4126 \\ \hline
% w/o \({L}_{\textit{mask}}\) & 23.16 & 0.7524 & 0.6513 & 0.4353 \\ \hline
% w/o \({L}_{\textit{adv}}\) & 24.18 & 0.8124 & 0.7318 & 0.3813 \\ \hline
% w/o \({L}_{\textit{cyc}}\)  & 23.83 & 0.7916 & 0.7132 & 0.3975 \\ \hline
% Ours      & \textbf{26.42} & \textbf{0.8648} & \textbf{0.8213} & \textbf{0.3511} \\ \hline
% \end{tabular}
% \end{table}
\begin{table}[h]
\caption{Ablation study of different loss functions.}\label{tab:diff_xiaorong}
\begin{tabular*}{\textwidth}{@{\extracolsep\fill}lcccc}
\toprule
Loss Function Setting & PSNR $\uparrow$ & SSIM $\uparrow$ & $CR$ $\uparrow$ & $S_{\text{smooth}}$ $\downarrow$ \\
\midrule
w/o \({L}_{\textit{diff}}\) & 23.65 & 0.7642 & 0.6942 & 0.4126 \\
w/o \({L}_{\textit{mask}}\) & 23.16 & 0.7524 & 0.6513 & 0.4353 \\
w/o \({L}_{\textit{adv}}\)  & 24.18 & 0.8124 & 0.7318 & 0.3813 \\
w/o \({L}_{\textit{cyc}}\)  & 23.83 & 0.7916 & 0.7132 & 0.3975 \\
Ours                        & \textbf{26.42} & \textbf{0.8648} & \textbf{0.8213} & \textbf{0.3511} \\
\botrule
\end{tabular*}
\vspace{-0.6cm}
\end{table}

\section{Discussion}
Our research on 2D medical angiography utilizes the parametric vascular model from Section~\ref{pvm} for precise guidance. However, this network faces several challenges.

Firstly, our 2D parametric vascular model is highly effective, as blood vessels appear continuous under X-rays. However, it cannot be directly applied to 3D volumes in CTA or MRA, necessitating the design of a 3D parametric vascular model.

Secondly, our method achieves better vascular masks under the guidance of the Parametric Vascular Model (PVM). The PVM is an essential component of our network structure. In Section~\ref{vascularfitting} and Section~\ref{diff_bezier}, we verify the fitting effects of different types of parametric vascular models. Additionally, we conducted further experiments to verify the impact of omitting the parametric vascular model on the entire network structure. These experiments show that if the parametric vascular model is missing, the network lacks more accurate semantic information guidance, making it difficult to synthesize high-quality angiographic X-rays.

\section{Conclusion}
We propose a self-supervised model for angiographic X-ray synthesis. This method employs an adversarial diffusion model in a fully unsupervised manner. Our method surpasses traditional pixel-based techniques by integrating vessel geometric features and employing adversarial learning to convert non-angiographic X-rays into angiographic X-rays. We contribute a new dataset SynthXCA of paired angiographic X-rays and non-angiographic X-rays with corresponding masks. Such paired data are usually inaccessible in real scenarios since vascular mask annotations require a lot of manual work from professionals. 

\textbf{Limitations. The geometric imaging effect of angiography is affected by scanning methods and varies across tissues and organs. While our adversarial diffusion-based synthesis method shows promising results, further research is required to enhance its generalizability and robustness under diverse imaging conditions and anatomical variations.}

\section*{Acknowledgements}
This work is supported in part by the NSFC (62325211, 62132021, 62372457), the Major Program of Xiangjiang Laboratory (23XJ01009), Young Elite Scientists Sponsorship Program by CAST (2023QNRC001), the Natural Science Foundation of Hunan Province of China (2022RC1104) and the NUDT Research Grants (ZK22-52).

% \section*{Declarations}

% Conflict of interest The authors declare that they have no known competing financial interests or personal relationships that could have appeared to influence the work reported in this paper.
% \begin{appendices}

% \end{appendices}

\noindent
% The input format for the above table is as follows:

%%=============================================%%
%% For presentation purpose, we have included  %%
%% \bigskip command. Please ignore this.       %%
%%=============================================%%

% \bibliography{sn-bibliography}% common bib file
%% if required, the content of .bbl file can be included here once bbl is generated
%%\input sn-article.bbl

% 参考文献表号从[1]改为1.
\makeatletter
\renewcommand\@biblabel[1]{#1.}
\makeatother

\bibliography{sn-bibliography}

% \bibliographystyle{unsrt}
% \bibliography{bib/references}

\end{document}